\documentclass[11pt]{article}
\usepackage[utf8]{inputenc}
\usepackage{cite}
\usepackage{listing}
\usepackage{listings}
\usepackage{graphicx}
\usepackage{longtable}
\usepackage{latexsym}
\usepackage{booktabs}
\usepackage{array}
\usepackage{longtable}
\usepackage{lipsum}
\usepackage{chngpage}
\usepackage{setspace}
\usepackage{xcolor}
\usepackage[format=plain,font=small,labelfont=bf]{caption}
\usepackage{float}
\usepackage{framed}
\usepackage{epsfig}
\usepackage{tabularx}
\usepackage{floatflt}
\usepackage{wrapfig}
\usepackage{makeidx}
\usepackage{multirow}
\usepackage[top=2.0cm, bottom=2.0cm, left=2.0cm, right=2.0cm]{geometry}
\usepackage[export]{adjustbox}
\usepackage[english]{babel}
\usepackage[section]{placeins}
\usepackage[toc,page]{appendix}
\usepackage[affil-it]{authblk}
\graphicspath{{figures/}}

\begin{document}

\title{\Large \bf Background Estimation Studies for Positron Double Beta Decay}
\date{}

\renewcommand\Authfont{\fontsize{12}{13}\selectfont}
\renewcommand\Affilfont{\fontsize{10}{9.8}\itshape}

\author[1]{Swati Thakur}
\author[2,3]{A. Mazumdar}
\author[2,3]{R. Shah} 
\author[2,3]{V. Vatsa} 
\author[4]{V. Nanal \thanks{nanal@tifr.res.in}}
\author[4]{M.S. Pose} 
\author[1]{Pushpendra P. Singh}
\author[1]{P.K. Raina}
\author[1]{R.G. Pillay}

\affil[1]{Department of Physics, Indian Institute of Technology Ropar, Rupnagar Punjab - 140001, India}
\affil[2]{India-based Neutrino Observatory, Tata Institute of Fundamental Research, Mumbai - 400005, India}
\affil[3]{Homi Bhabha National Institute, Anushaktinagar, Mumbai - 400094, India}
\affil[4]{Department of Nuclear and Atomic Physics, Tata Institute of Fundamental Research, Mumbai - 400005, India}

\maketitle

\section*{Abstract}
The study of neutrinoless double beta decay has attracted much attention as it can provide valuable information about the mass and the nature of the neutrino. The double beta decay (DBD) itself is also of interest in nuclear physics. While DBD has been observed in about a dozen nuclei, the positron double beta decay ($\beta^{+}\beta^{+}$/EC-$\beta^{+}$) continues to be an elusive. An important signature for $\beta^{+}\beta^{+}$ decay is the simultaneous emission of four 511\,keV gamma rays and the coincident detection of these gamma rays can improve the measurement sensitivity.  
This paper presents an estimation of sensitivity for EC-$\beta^{+}$ and $\beta^{+}\beta^{+}$ employing coincidence measurement with two high purity Ge (HPGe) detectors. Simulations for coincident detection efficiency (${\epsilon}_c$) of 511\,keV gamma rays with two HPGe detectors have been carried out using GEANT4 for different source geometries to optimize the mass efficiency product (M${\epsilon}_c$).  
The source of size $55\,mm \times 55\,mm \times 5\,mm$ (thickness) sandwiched between the front faces of the detectors were found to be optimal for 2 pairs of 511\,keV gamma rays in the present detector setup.  The coincident background is estimated at the sea level with moderate Pb shielding. With this setup, the sensitivity for T$_{1/2}$ measurement of EC-$\beta^{+}$ in $^{112}$Sn and $\beta^{+}\beta^{+}$ in $^{106}$Cd is estimated to be $\sim$\,10$^{19}$ - 10$^{20}$\,y for 1\,y of measurement time. 

\section{Introduction}
Recent neutrino oscillation experiments have boosted the worldwide interest in the search for neutrinoless double beta decay (NDBD) with increased sensitivity~\cite{dolinski2019}. The neutrinoless double beta decay provides a unique probe to study the mass and nature of the neutrino. Double beta decay (DBD) is a second-order allowed process in the Standard Model and $\beta^{-}\beta^{-}$ decay mode has been detected in several nuclei. 
However, positron decay modes are hindered due to lower effective Q value~\cite{belli2021} and experimental search for $\beta^+\beta^+$/EC-$\beta^+$ continues to be elusive.
An important experimental signature of the decay modes involving positrons like EC-$\beta^{+}$ and $\beta^{+}\beta^{+}$ is the simultaneous emission of pair(s) of 511\,keV gamma rays.   
Hence, the coincident detection of 511\,keV gamma rays can significantly improve the measurement sensitivity. However, since 511\,keV gamma rays can also originate from many other processes, it is important to understand and discriminate against the background originating from trace impurities in the source, detector and surrounding materials and as well as from cosmic muon induced reactions. 
The present work aims to estimate the sensitivity of the half-life measurement for positron DBD modes using a coincidence setup. In the first part, simulations are done to optimize the source-detector configuration to maximize the mass efficiency product (M${\epsilon}_c$). In the second part, background measurements are carried out using a coincidence setup of two high purity Ge (HPGe) detectors ($\sim33\%$) with moderate lead shielding. Both ambient background and background with natural tin ($^{nat}$Sn) sample of mass $\sim$\,40\,g were measured.  The background in the coincidence setup, especially in the region of interest around 511\,keV is compared with that from the low background counting setup TiLES~\cite{neha2014}. The sensitivity for T$_{1/2}$ measurement of EC-$\beta^{+}$ in $^{112}$Sn and $\beta^{+}\beta^{+}$ in $^{106}$Cd is estimated using the measured background and simulated coincidence efficiency. Measures for improvement in the background are also discussed. 

\begin{figure}[h!]
\centering
\includegraphics[width=7.6cm,height=5.0cm, trim=180 400 20 150, clip]{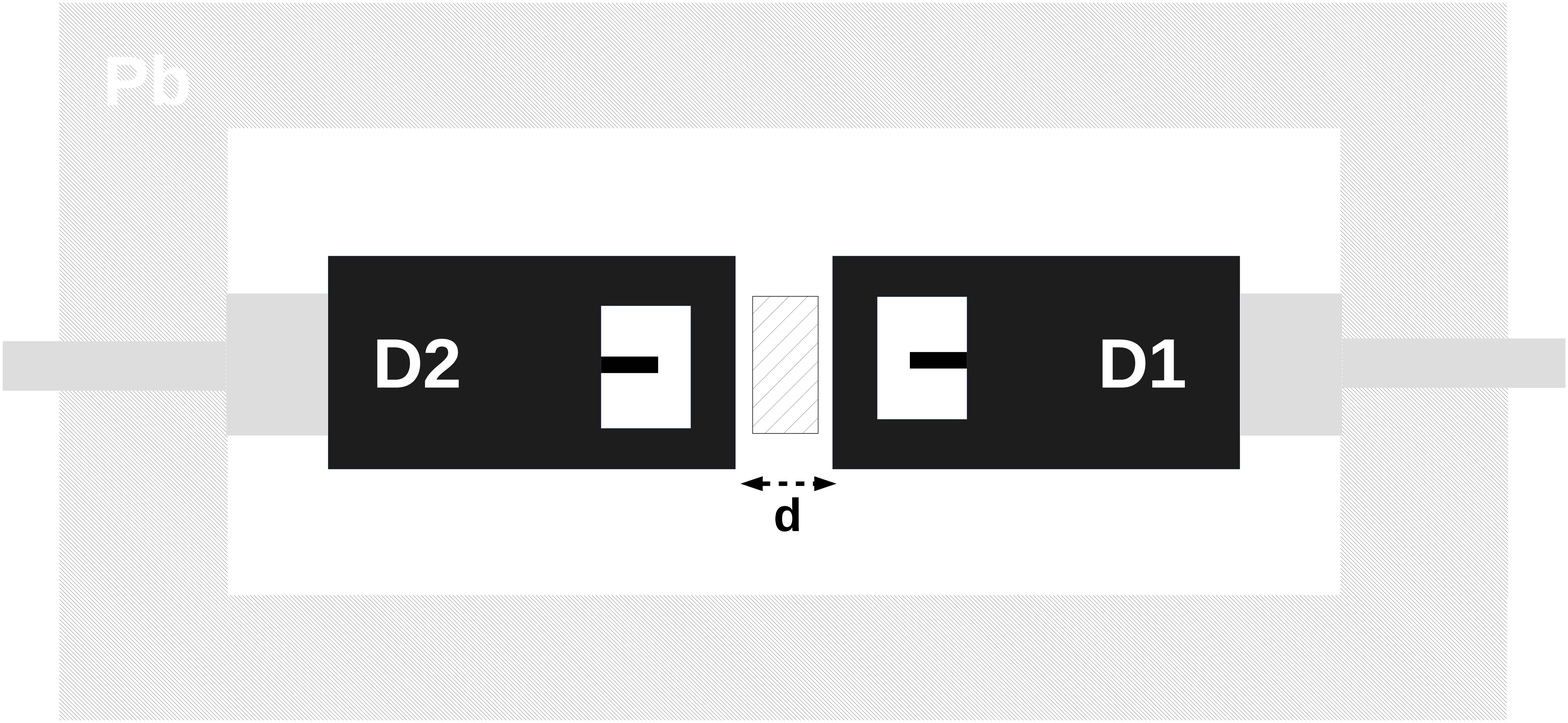}
\includegraphics[width=6.0cm,height=5.0cm, trim=0 0 0 0, clip]{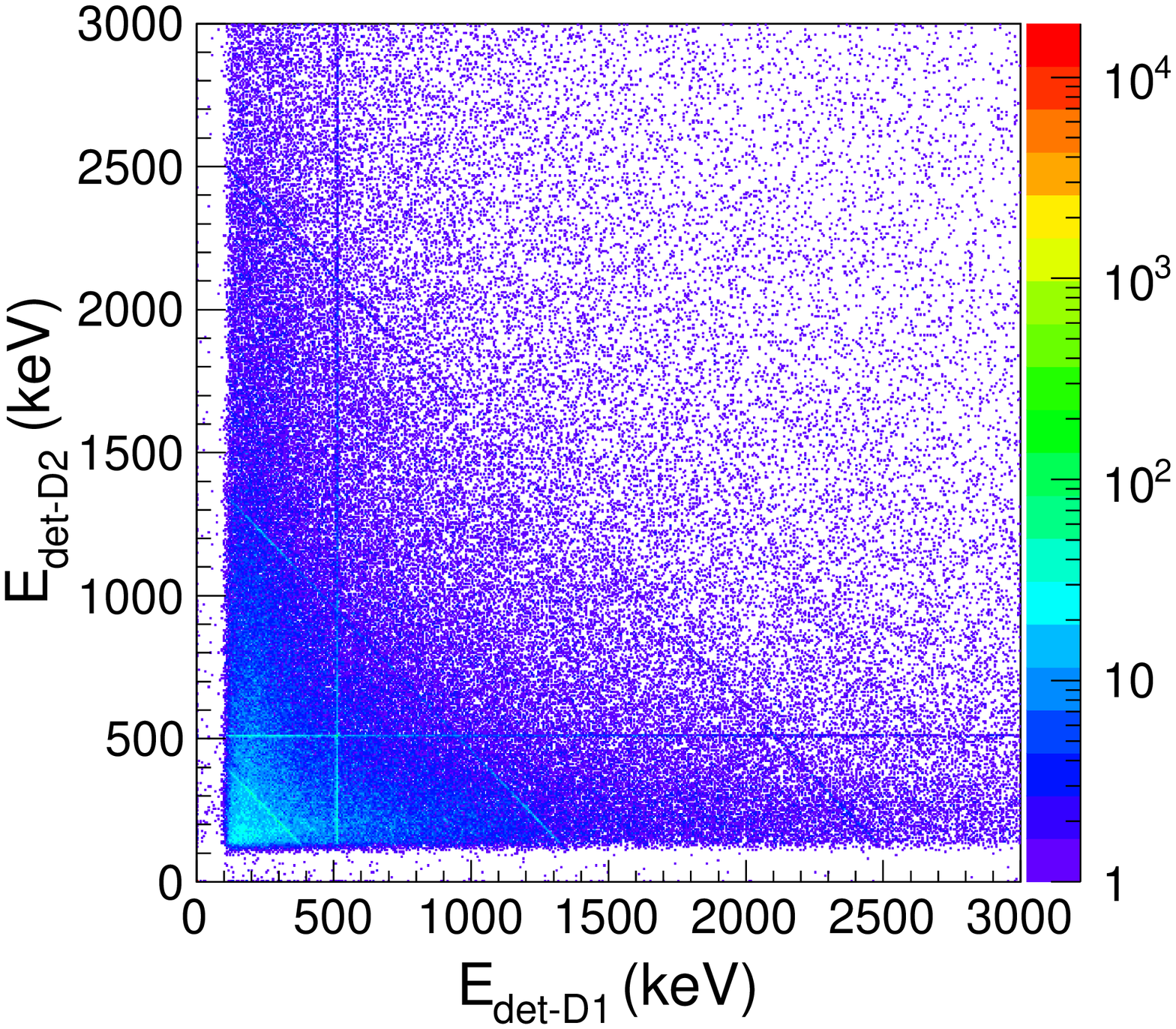}
\caption{\label{fig:setup} a) A schematic view of D1-D2 coincidence setup and b) Coincident energy spectra of D1-D2 for ambient background ($t\,=\,27.0$\,d).}
\end{figure}

\section{Experimental details}
For background measurements in coincidence, a simple setup with two identical low background cryo-cooled HPGe detectors is made at TIFR, Mumbai. Both detectors D1 and D2 (Ortec make) have carbon fiber housing with 0.9\,mm thick front window, and $\sim$\,33\,\% relative efficiency each. Detectors are mounted in a close geometry with a face-to-face distance of about 2\,cm as shown in Figure~\ref{fig:setup}(a). The detailed measurements were carried out with detector D1 using mono-energetic point+volume sources, and an optimized geometrical model of the detector was obtained~\cite{gupta2018}. The same geometry is adopted for both D1 and D2 in the present simulations. This two detector setup is surrounded by a passive shield of 5\,cm thick low activity lead ($<$\,0.3\,Bq/kg) in all directions. Additionally, 5~cm thick lead ($<21$\,Bq/kg) is added on both the sides. The setup also has a provision for cosmic muon veto in the future.

Data were acquired using CAEN DT6724 digitizer (14-bit, 100 MS/s) and recorded separately (time stamp and energy) for each detector on an event by event basis. The coincident spectra were generated using offline analysis.
The dead time was monitored with a standard 10\,Hz pulser, and was found to be negligible ($<$ 0.1\,\%). The resolution of the detector was $\sim$\,3\,keV at 1332\,keV.
The ambient background was recorded for $\sim 27$\,d, at different times over the period of about 10 months. No measurable drifts were observed in the data.
The $^{nat}$Sn sample of mass 38.8990\,g, in the form of  granules (7N purity, Alfa aesar) with an approximate overall size  $30\,mm \times 39\,mm \times 6.5\,mm$,  was counted in close geometry for $t\,=\,77.8$\,d. Figure~\ref{fig:setup}(b) shows a typical 2-dimensional plot of E$_{D1}$ vs E$_{D2}$ for the ambient background. Correlated 511\,keV lines and high energy background lines (1460, 2615\,keV) are clearly visible.

The coincidence efficiency ($\epsilon_c$) of D1-D2 detectors for $^{nat}$Sn and $^{nat}$Cd foils was obtained using a GEANT4~\cite{geant} simulation program. A total of $10^6$ events of 2 (1) correlated pairs of 511\,keV gamma rays were generated at a given vertex for $\beta^{+}\beta^{+}$ (EC-$\beta^+$). 
In rare decay search experiments like $\beta^+\beta^+$, the mass efficiency product (M$\epsilon_{c}$) needs to be maximized. Hence, the source size and mounting geometry were optimized for the present detector setup. Details of simulation and source optimization procedure are given in  Ref.~\cite{thakur2021simulation}. 
It was found that 55\,mm$ \times $55\,mm source of 10\,mm thickness yields maximum M$\epsilon_{c}$. However, with a 5\,mm thick source, the detectors can be moved closer (d\,= 7\,mm) and consequently decrease in M$\epsilon_{c}$ is only $\sim$\,20\,\% even if M is reduced by 50\,\%. This is preferable as the inherent background scales with the mass of the source. Hence, the optimal source geometry is chosen to be $55\,mm \times 55\,mm \times 5\,mm$ foil sandwiched between front faces of detectors, with separation between detectors d\,= 7\,mm. The simulated values of $\epsilon_{c}$ are given in Table~\ref{tab:table1}.

\section{Data analysis and results}
\begin{figure}[h!]
\begin{center}
\includegraphics[width=11.2cm,height=7.5cm, trim=0 0 0 0, clip]{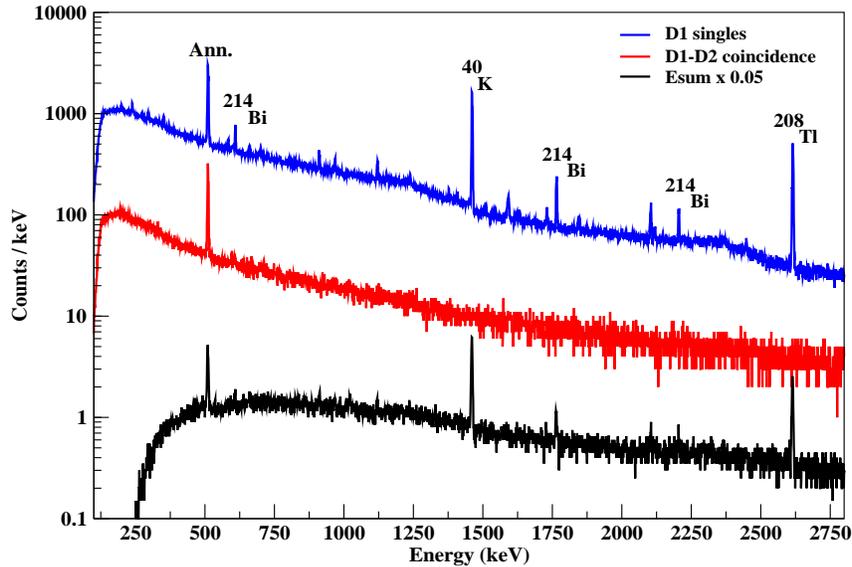}
\end{center}
\caption{\label{fig:bkg_comp} A comparison of the ambient background in D1 - singles and coincidence with D2 (i.e. E$_{D2}$\,$>$\,0). Sum energy spectrum E$_{sum}$\,=\,E$_{D1}$+E$_{D2}$, scaled by an arbitrary factor of 0.05 for better visibility, is also shown for comparison. All spectra have been time normalized to t$\,=\,$7\,d.}
\end{figure}

The coincidence was performed using an offline C++ based algorithms developed in ROOT~\cite{root}. The coincidence time window was set to $\pm$\,1\,$\mu$s to ensure that all coincident events are collected, and the output was written in a ROOT Tree. The data were analyzed using LAMPS~\cite{lamps}. Figure~\ref{fig:bkg_comp} shows a comparison of the ambient background of D1 in singles and  coincidence. It is evident that the coincidence yields significant improvement in the background. In fact, only 511\,keV peak survives in the background, albeit with much reduced intensity. The sum energy spectrum (E$_{sum}$\,=\,E$_{D1}$+E$_{D2}$) is also shown for comparison. It can be seen that while high energy gamma rays like 1460\,keV ($^{40}$K) and 2615\,keV ($^{208}$Tl, originating from $^{232}$Th) are visible in sum energy, the overall background is reduced by an order of magnitude.

For both ambient background and $^{nat}$Sn sample, analysis to extract counts in the region of interest (ROI) near 511\,keV was done in an identical manner. The chance correction from the time spectrum was found to be negligible. The coincident 511-511\,keV events were suitably corrected for underlying Compton chance coincidence.
The prompt gated D1 spectrum was generated for the photopeak region (511 $\pm$ 5\,keV) in D2, while the chance gated spectrum was generated from 5\,keV window on the left and right of the photopeak in D2. The observed counts in the ROI for ambient background (i.e., without the sample) are 3 $\pm$ 1\,cts/d, which is equivalent to 1271 $\pm$ 297\,cts/y. 
It should be mentioned that the observed singles count rate for 511\,keV gamma ray in low background setup TiLES~\cite{neha2014}, is 680 $\pm$ 20\,cts/d, which clearly emphasizes improvement with coincident detection. However, higher statistics will be required to study 1022 - 1022\,keV correlations.
In presence of $^{nat}$Sn ($\sim$\,40\,g), the background in the ROI was enhanced to 1919 $\pm$ 211\,cts/y, giving the excess of 2 $\pm$ 1\,cts/(keV.g.y) with $^{nat}$Sn. 

In the absence of a positive signal, based on the present background, a lower limit on the half-life can be estimated as:
\begin {equation}\label{eq:1}
 T_{1/2} > \frac{ln2*N_A*\epsilon_c*a}{w*k_{CL}} \sqrt{\frac{M*t}{N_{bkg}*\Delta E}}
\end {equation} 
where $N_{A}$ is Avogadro’s number, $\epsilon_c$ is the coincidence detection efficiency of a gamma ray, $a$ is the isotopic abundance of sample, $W$ is the molar mass of sample, $M$ is the sample mass, $t$ is the time of measurement (in y), $N_{bkg}$ is the background index (in cts/(keV.g.y)), $\Delta E$ is the energy window defining the signal region of D1-D2 setup and $k_{CL}$ is the number of standard deviations corresponding to a given confidence interval (C.L.).
If one considers only the background from the source,  N$_{bkg}$ is estimated to be 2 $\pm$ 1\,cts/(keV.g.y) from the tin data. As mentioned in previous section, the $\epsilon_c$ is obtained from simulations for an optimal source geometry ($55\,mm\times 55\,mm\times 5\,mm$). Using the simulated $\epsilon_c$, the  $T_{1/2}$ sensitivity of the present D1-D2 coincidence setup for EC-$\beta^{+}$ in $^{112}$Sn and $\beta^{+}\beta^{+}$ in $^{106}$Cd has been estimated for different enrichment fractions and listed in Table~\ref{tab:table1}. 

\begin{table} [h!]
\centering
\caption{\label{tab:table1} The projected sensitivity for the half-life (T$_{1/2}$) of the present D1-D2 coincidence setup for $^{112}$Sn (EC-$\beta^{+}$) and  $^{106}$Cd ($\beta^{+}\beta^{+}$) for $t_{data}\,=\,$\,1\,y. The isotopic abundance ($a$), total mass ($M_{0}$) and coincidence efficiency ($\epsilon_c$) are also listed.}
\begin{tabular}{lccccc}
\hline
\rule{0pt}{2.5ex}
Source & $a$ & $\epsilon_c$ & $M_{0}$  & T$_{1/2}$ (68\% C.L.) & T$_{1/2}$ (90\% C.L.) \\
    & (\%)   & (\%)         & (g)        & (y)   & (y) \\
\hline
\rule{0pt}{2.5ex}
$^{106}$Cd & 50 & 1.04        & 130        & 5.5$\times$10$^{19}$  & 3.4$\times$10$^{19}$ \\
$^{106}$Cd & 90 & 1.04        & 130        & 1.0$\times$10$^{20}$  & 6.2$\times$10$^{19}$ \\
\hline
$^{112}$Sn & 50 & 0.64        & 110        & 2.9$\times$10$^{19}$  & 1.8$\times$10$^{19}$ \\
$^{112}$Sn & 90 & 0.64        & 110        & 5.4$\times$10$^{19}$  & 3.3$\times$10$^{19}$ \\
\hline
\end{tabular}
\end{table}

The ambient background has contributions from trace radioactive impurities, natural radioactive chains and cosmic muons as well as muon induced reactions. The external background can be minimized with suitable shielding. A moderate rock cover of $\sim$\,500\,m would suppress the muon flux by $\sim$\,4 orders of magnitude~\cite{Jaduguda}. Thus, the inherent background from the source (trace impurities, neutron induced reactions) will be a limiting factor and hence the same has been used in the present estimation. Nevertheless, it is important to reduce overall background. The use of larger detectors with an annular anti-compton shield would also improve the coincidence efficiency and reduce background~\cite{finch2016}. In the present setup, it is proposed to augment the shielding by adding an active veto for muon and increasing passive shield thickness.  

The present best limits for $T_{1/2}^{\beta^{+}\beta^{+}}$ ($^{106}$Cd) and $T_{1/2}^{EC-\beta^{+}}$ ($^{112}$Sn) are $2.3 \times 10^{21}$\,y~\cite{belli2016} and  $9.7 \times 10^{19}$\,y~\cite{barabash2011}, respectively. From Table~\ref{tab:table1}, it can be seen that an improvement in the background index by about a factor of 5, will be suitable to yield an improved limit for $^{112}$Sn, while for $^{106}$Cd further measures to improve signal to noise ratio are essential.
 
\section{Conclusion}
The feasibility study of positron double beta decay modes is presented for $^{112}$Sn (EC-$\beta^{+}$) and $^{106}$Cd ($\beta^{+}\beta^{+}$) using a coincidence setup of 2 HPGe detectors.
The coincident detection efficiency of 511\,keV gamma rays for source foil sandwiched between the detectors has been estimated using GEANT4. 
The source of size $55\,mm \times 55\,mm \times 5\,mm $ (thickness) was found to be optimal for 2 pairs of  511\,keV gamma rays.  The ambient background of the 2 detector setup with moderate Pb shielding is measured in coincidence mode at sea level. The coincident detection of 511\,keV pair yields a significant improvement in the background in the region of interest. From background measurements with $\sim$\,40\,g of $^{nat}$Sn, the sensitivity for  $T_{1/2}^{\beta^{+}\beta^{+}}$  ($^{106}$Cd) and $T_{1/2}^{EC-\beta^{+}}$ ($^{112}$Sn) are estimated to be $\sim$\,10$^{19}$ - 10$^{20}$\,y for 1\,y of measurement time with enriched samples. 
Thus, coincidence measurements with the present two HPGe detector setup at moderate depth can be used to probe EC-$\beta^{+}$/ $\beta^{+}\beta^{+}$.
 
\section*{Acknowledgement}
We thank  Mr.~S.~Mallikarjunachary and Mr.~K.V.~Divekar for the assistance with measurements. This work is supported by the Department of Atomic Energy, Government of India (GoI), under Project No. RTI4002. Swati Thakur acknowledges the Ministry of Education, GoI, for PhD research fellowship.


\begin{thebibliography}{100}
\bibitem{dolinski2019} M. J. Dolinski {\it et. al.} Annu. Rev. Nucl. Part. Sci. {\bf 69}, 219 (2019).
\bibitem{belli2021} P. Belli {\it et. al.} Particles {\bf 4}, 241 (2021).
\bibitem{neha2014} N. Dokania {\it et. al.} Nucl. Instrum. Meth. A {\bf 745}, 119 (2014).
\bibitem{gupta2018} G. Gupta {\it et. al.} DAE-BRNS Symp. Nucl. Phys. {\bf 63}, 1142 (2018).
\bibitem{geant} S. Agostinelli {\it et. al.} Nucl. Instrum. Meth. A {\bf 506}, 250 (2003).
\bibitem{thakur2021simulation} S. Thakur {\it et. al.} arXiv 2110.02171 (2021).
\bibitem{root} R. Brun {\it et. al.} Nucl. Instrum. Meth. A {\bf 389}, 81 (1997).
\bibitem{lamps}https://www.tifr.res.in/$\sim$pell/lamps.html
\bibitem{Jaduguda} M. K. Sharan {\it et. al.} Nucl. Instrum. Meth. A {\bf 994}, 165083 (2021).
\bibitem{finch2016} S.W. Finch {\it et. al.} Nucl. Instrum. Meth. A {\bf 806}, 70 (2016).
\bibitem{belli2016} P. Belli {\it et. al.} Phys. Rev. C {\bf 93}, 045502 (2016).
\bibitem{barabash2011} A. S. Barabash {\it et. al.} Phys. Rev. C {\bf 83}, 045503 (2011).
\end{thebibliography}
\end{document}